\documentclass[preprint2]{aastex63}
\submitjournal{AAS Journals}

\begin{document}

\title{A search for Planet Nine using the Zwicky Transient Facility public archive}

\correspondingauthor{Michael Brown}
\email{mbrown@caltech.edu}

\author[0000-0002-8255-0545]{Michael E. Brown}
\affiliation{Division of Geological and Planetary Sciences\\
California Institute of Technology\\
Pasadena, CA 9125, USA}

\author[0000-0002-7094-7908]{Konstantin Batygin}
\affiliation{Division of Geological and Planetary Sciences\\
California Institute of Technology\\
Pasadena, CA 91125, USA}

\begin{abstract}
Recent estimates of the characteristics of Planet Nine have
suggested that it could be closer than originally assumed. 
Such a Planet Nine
would also be brighter than originally assumed, 
suggesting the 
possibility that it has already been observed in wide-field moderate-depth
surveys. We search for Planet Nine in the Zwicky Transient Facility public
archive and find no candidates. Using known asteroids to
calculate the magnitude limit of the survey, we find that we should
have detected Planet Nine throughout most of the northern 
portion of its predicted orbit -- including within the 
galactic plane --
to a 95\% detection efficiency
of approximately $V=20.5$. To aid in understanding detection
limits for this and future analyses, we present a full-sky synthetic
Planet Nine population drawn from a statistical sampling 
of predicted Planet Nine orbits. We use this reference population to estimate
that this survey rules out $56\%$ of predicted Planet Nine phase space,
and we demonstrate how future analyses can use the same synthetic population
to continue to constrain the amount of parameter space effectively
searched for Planet Nine.
\end{abstract}

\section{Introduction}
The unexpected alignment of both the longitudes of perihelion and the orbital
poles of distant detatched
Kuiper belt objects (KBOs) with semimajor axes beyond
150 AU suggests the existence of a giant planet well beyond
the Kuiper belt \citep{2016AJ....151...22B}. 
The preliminary estimates of the orbital parameters of this
planet -- which we will call Planet Nine here --
suggested that the planet could have a semimajor of 700 AU or greater and
a visual magnitude of 24 or fainter.
Several observational searches are
currently underway for this hypothesized planet
using large telescopes capable of reaching such faint distant objects.

Recently, \citet[][hereafter BB21]{brownp9} developed a method
to use the full set of detections of distant 
detached KBOs, along with estimates of the biases in
these detections and a large suite of numerical simulations, 
to make a statistical model
for the orbital parameters of Planet Nine.
The posterior distribution for the predicted 
distance to Planet Nine is broad, but
one important outcome of the analysis is
the realization that Planet Nine could possibly
be both closer and less
massive than originally assumed. 
Even though a less massive planet will 
likely be slightly smaller,  the closer distance to the
planet can make it as much as 
2 magnitudes brighter than the original
estimates. Planet Nine may not require dedicated searches on large
telescopes but, instead, may have been already imaged in one of the
increasingly large numbers of wide field surveys completed or underway
to date.

Here, 
we search the first 3 years 
of the public Zwicky Transient Facility (ZTF)
archives for Planet Nine using a new orbit-linking algorithm that is
efficient at searching sparsely sampled data across multiple-opposition. 
To understand the effectiveness
of our search
we develop a method to calibrate the efficiency of the ZTF
observations
using both detections and non-detections
of known asteroids within the ZTF survey. 
The survey
method and calibration are applicable to other large-scale transient
surveys currently underway.

\section{ZTF data}
Our orbit-linking method and calibration are agnostic to how and where the
data are acquired. Instead, our self-calibration accounts for the
characteristics of the survey depth and cadence with
no pre-knowledge. Nonetheless, we briefly introduce the key characteristics
of the ZTF public survey here.
The ZTF is a time-domain survey run on the 
Samuel Oschin 48” Schmidt telescope at Palomar Observatory. The CCD 
camera covers a 47 deg$^2$ field of view. 
The public survey nominally
covers the full sky visible north of -30$^\circ$ declination once every
three nights. 
In addition, a strip approximately 10$^{\circ}$ wide
centered on the galactic plane is imaged nightly. 
The 15 second exposures of the survey reach a $5\sigma$ 
reported 
depth of $g=20.8$ and $r=20.6$, though significant variations occur \citep{ztf_overview}. { Astometric registration of the frame to GAIA is
good to 30 milliarcseconds, and the faintest sources cataloged have uncertainties
of $\sim$ 1 arcsecond \citep{2019PASP..131a8003M}.}
The archive
begins with data on 1 June 2018 and continues to be updated nightly. 
We restrict
our analysis to data from 1 June 2018 until 7 May 2020
and to the specific region on the sky predicted for
the location of Planet Nine by BB21, 
The ZTF data cover essentially every location along the predicted
path of Planet Nine many 
times over on many separate dates with the exception of
the furthest southern portion of the orbit. 

The ZTF public archive contains only nightly transients, that is,
detections on the image made after a reference template has been
subtracted \citep{2019PASP..131a8003M}. Such transients are ideal for detecting moving objects
in the solar system. 
Every morning, we download the full set of nightly transients from
the archive page\footnote{\url{http://ztf.uw.edu/alerts/public}}. 
Many real astrophysical transients
will repeat sporadically or stay bright for multiple nights;
we thus discard any transients that appear more than once at 
a single location.
As of 1 June 2018, the Planet Nine search area of the
ZTF archive contains $\sim$13 million detections that
fit this criteria, including 8.6 million within 2 arcseconds of known
solar system objects (discussed further below). 
Nearly every other single-night transient is
a false positive of some sort, but unknown bright solar system objects
will also appear -- possibly many times -- in this catalog. Our task is to
discard the many false positives and 
link any real objects onto Keplerian orbits to identify their existence.

\

\section{Detection limits}
Before searching for Planet Nine itself, we develop a method
to determine detection limits in the ZTF data.
Little information on coverage, filling-factor,
sky conditions, or other parameters that would be required to
compute real search limits is available in the 
nightly public archive. We instead
develop a method using known asteroids to self-calibrate the survey.

Every ZTF image -- particularly near the ecliptic -- contains 
many detections of asteroids, 
each of which will appear in the single-epoch
transient catalog that we create. Numbered asteroids have sufficiently
well known orbits that they should appear at nearly precisely their
predicted position. These asteroids also have predicted magnitudes.
Unfortunately, the predicted magnitudes are not as reliable
as the predicted positions. The predicted magnitudes do not account
for light curves, colors, or individual phase functions. To determine
how much the true magnitudes differ from
the predicted magnitudes we examine 100,000
labeled asteroids and find that, accounting for offsets between $g$ and $r$
and the predictions, which are in $V$, the predictions have a root-mean-square (RMS)
deviation of 0.2 magnitudes with respect to the observations. No systematic offsets
are seen.

We will thus use these asteroids as known transient
point sources of known magnitude to calibrate the efficiency of the survey, with the caveat
that the calibration is uncertain to a few tenths of a magnitude.
The benefit of using these asteroids is that, { owing to the large
number of asteroids on each image taken}, they account for
the efficiency of all aspects of the survey, from observing
conditions { at the moment of the image}, 
to { which regions of the sky are surveyed each night}, 
to detector geometry, to 
transient processing pipeline.

We use JPL Horizons\footnote{\url{http://ssd.jpl.nasa.gov}} to predict position and magnitudes
at 1 day intervals across the 3 years of the survey of every
numbered asteroid. We then interpolate to the time of each image and
look for transients within 2 arcseconds of each predicted asteroid
position. 
We find that most 
of the asteroid associations are labeled in the ZTF archive, but
a small number are not. { Given the astrometric precision of
	ZTF and the well-known orbits of numbered asteroids, most
	of the detections are within an arcsecond of their predicted
position. }

For each night in the 3 year period, we track each numbered
asteroid that appears in the transient catalog and also each that
could have appeared, but does not. We divide the sky into $\sim$1.8 square degree equal-area regions using an NSIDE=32
HEALPix grid
\footnote{\url{https://healpix.jpl.nasa.gov/html/idl.htm}},
and we grid each detection and non-detection
into 0.25 magnitude bins for each night of the 3 years. Note that
we exclusively use the predicted magnitude for both detected and 
undetected objects. While ZTF observes in multiple wavelength bands,
we make no attempt at correcting for color; this procedure is equivalent
to implicitly assuming that Planet Nine has the same color as the
average asteroid. Within the limits of this calibration, this assumption
should be adequate.
For each night
and at each position in the sky, we are thus able to estimate the probability
that a moving object of a particular predicted magnitude is detected any night of the 3 year survey.

As an initial estimate of the Planet Nine search efficiency,
we can ask the question: what is the probability that an object of a given magnitude
would be detected $n$ or more times at a particular location? 
Here, $n$ is the number of detections that we require to successfully
detect Planet Nine. Smaller values of $n$ will make it more
likely that enough detections of Planet Nine will have occurred,
while increasing the probability of false linkages. Higher values of
$n$ will lower our detection efficiency but likewise lower
the false positive rate. In addition, higher values of $n$
require less processing time for our detection algorithm, described below.
Below, we find that $n\sim 7$ provides
an attractive trade-off between these competing factors.
For this initial detection limit estimation, 
we thus
define 
our magnitude limit as the brightness at which there is a 95\% or
greater probability that an object at that magnitude would
have been detected 7 or more times. To determine this magnitude limit, 
we simulate 1000 observations in each HEALPix at each point in the magnitude
grid on each night of the survey and determine the faintest object
that is detected 7 or more times 95\% of the time. Our survey
limit map is shown in Fig 1.
\begin{figure*}
\plotone{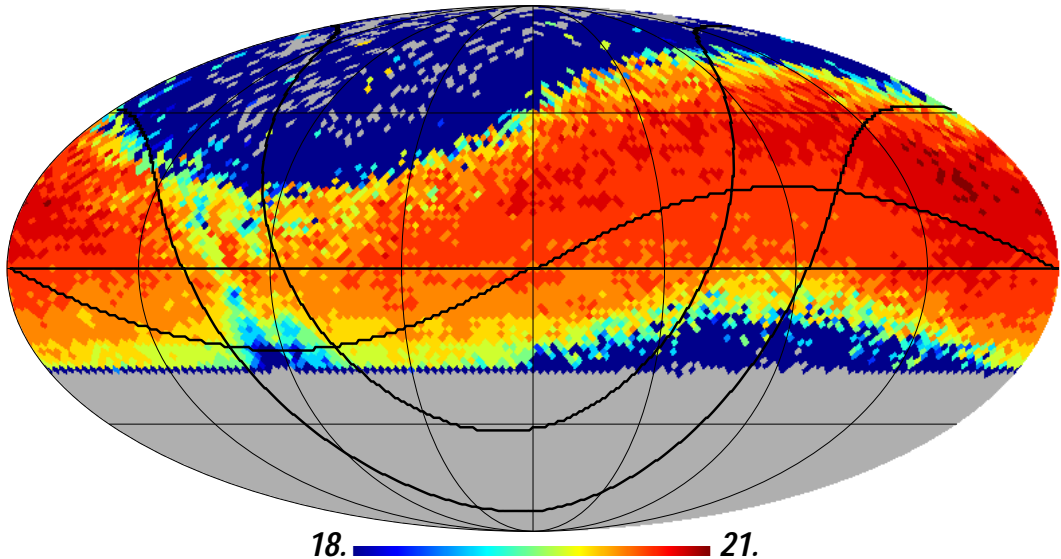}
\caption{The V magnitude at which there is a 95\% or higher probability
that a moving object would be detected 7 or more times in the 
first 3 years of the ZTF public survey. The Mollweide equal-area projection
is centered at RA=180$^{\circ}$ and dec=0$^{\circ}$ with Right Ascension increasing towards the left. 
The path of the ecliptic and $\pm15^\circ$ of the
galactic plane are highlighted. For much of the sky a uniform limit
of $V\sim 20.5$ is attained, with slightly greater depth north of celestial equator.
Despite the high density of stationary stars, the northern galactic plane has no
worse depth than any other region of the sky. The higher density in the southern galactic
plane and the lower declinations make this region the least well-covered.
At high ecliptic latitudes the small
number of asteroids prevents accurate calculation of the limits.}
\end{figure*}

Several interesting features of the limit map are clear. First, 
the northern galactic plane is well covered. Even with the high density
of stationary sources in the plane causing confusion with transients,
the high cadence -- particularly within 10 degrees of the plane --
means that real moving objects have multiple opportunities to be detected
throughout the year and are eventually detected 7 or more times
as efficiently as outside of the galactic plane.
The southern galactic plane is less-well covered, owing to the 28 degree
latitude of Palomar observatory, but regions of good coverage exist.
The survey limit estimate is best where large numbers of asteroids are
predicted or observed. At higher ecliptic latitudes the method fails. 
In principle one could estimate an efficiency based on stellar
density and coverage, though we do not attempt that here.

In order to understand how strongly the number-of-detections-required
affects detection limit, we calculate a limit map for 14 required
detections. 
Throughout most of the survey region, the limits are worse
by no more than 0.25 magnitudes. The extremely large number of images obtained
across the whole northern sky by ZTF ensure that a sufficiently bright Planet Nine
will be detected many times in the survey.

\section{A Planet Nine reference population}
For calibration of our search algorithm, described below, and
for eventual understanding of its detection limits, it will
be helpful to define a Planet
Nine reference population and simulate its detection. 
Indeed, this reference population
is a useful tool beyond this analysis, as it will
allow any survey to help understand which parts of the 
Planet Nine parameter space it has ruled out and which others
have already been ruled out.

The Markov Chain Monte Carlo analysis of  BB21
created a posterior sample of orbital parameters for 49100
realizations of Planet Nine. 
{ 
For these realizations, 95\% of the population is within a swath of sky $\pm 12^\circ$ in declination from an orbit with an inclination of $16^\circ$ and an ascending node of $97^\circ$. The 1$\sigma$ bounds
on the population has a mass of $6.2^{+2.2}_{-1.3}$ Earth masses, a semimajor axis of $380^{+140}_{-80}$ AU, and
a perihelion distance of $300^{+85}_{-80}$ AU.}

We create our reference population of 100,000 potential
Planets Nine by randomly selecting from the sample parameters
100,000 times and picking a random value of mean anomaly, $M$, 
between 0 and 360$^\circ$. We use the same assumptions as
BB21 for the radius and albedo, that is
we assume a simple mass-diameter relationship of 
$r_9=(m_9/{3 \rm M}_{\rm earth})$ R$_{\rm earth}$ based
on fits to planets in this
radius and mass range by \citet{2013ApJ...772...74W}, and we assume
a full range of albedos from 0.2 -- half that of Neptune -- to 0.75,
the value predicted by modeling from
\citet{2016ApJ...824L..25F},
who find that all absorbers are condensed out of the
atmosphere and the planet should have a purely Rayleigh-scattering
albedo. 

This Planet Nine reference population contains a statistical
sampling of orbital
elements -- including a mean anomaly at a defined epoch -- 
and a mass, radius, and albedo for each member of our synthetic sample.
The sample, permanently archived at \url{https://data.caltech.edu/records/2098} \citep{brown_2021}, can be used to predict the position
and brightness of all members of 
the Planet Nine reference population for any future or past survey.
We use this sample to  predict
position and magnitude
for each night of the 3 year ZTF survey. We then use our 
asteroid-derived estimate of detection efficiency as a function of HEALPix, 
magnitude, and observing night, to determine the probability
that each synthetic object would be observed on any given night.
We select a uniform random number between 0 and 1 and, if the
number is lower the probability of detect of an object of that 
magnitude, we record that object as detected at that position on that night.

As we will discuss below, we will require the detection of
Planet Nine at least 7 times over our 3 year period. Based on
the cadence of observations and our calculated magnitude limits, 
56373 of the 100000 objects in the
Planet Nine reference population would have been detected
7 or more times in the ZTF survey. { The undetected members of the reference
are either too faint for the ZTF survey or south of the survey latitude 
limit}. A total of 42350 objects are detected
as many as 25 times. These reference population detections show that, if we 
can develop an algorithm capable of linking 7 or more
Planet Nine-like orbits, the ZTF survey will
be capable of ruling out more than 56\% of potential Planet Nine 
parameter space. 

\section{Orbit linking}
\citet{2015AJ....149...69B} developed a general method for linking single-epoch
transient detections across multiple nights onto outer solar system
orbits. The method 
involved calculating best-fit orbits to
every possible set of three transient detections 
across a single opposition using the geocentric linear
approximation of \citet{2000AJ....120.3323B}, which we refer to as the Bernstein
method. For triplets with sufficiently small
residuals in the linear approximation, full orbits were determined
and residuals were determined. Only orbits with residuals below
1 arcscond were retained. The large numbers of triplets
were eventually combined to find low-residual
quadruplet linkages and,
for particularly contaminated regions, to quintuplets. For the data
from the 8 years of the Catalina Realtime Transient Survey, this 
process led to detections of all bright Kuiper belt objects within
the survey field with zero false positives, but zero new detections.

While effective, this method does not scale well to the larger surveys
now becoming available. Recently, \citet{2018AJ....156..135H} developed an
alternative orbital approximation which greatly improves the
ability to link multiple epochs even across multiple opposition.
In the simplest version of the Holman method, objects are assumed
to be moving at a constant speed in circular motion
around the sun.
The observed position of an object is 
then transformed into
heliocentric, rather than geocentric, coordinates, using a range 
of assumed
distances and the known position of the Earth. In this coordinate system
objects move on great circles in heliocentric coordinates
at a constant speed when the correct distance is assumed, 
greatly simplifying linking (Figure 2). This method is increasingly efficient at larger
heliocentric distances where the constant distance approximation is generally closer to being
true over the orbital arc being fit. It is thus an attractive method
to try to link Planet Nine over multiple oppositions.
\begin{figure}
   \epsscale{1.3}
   \plotone{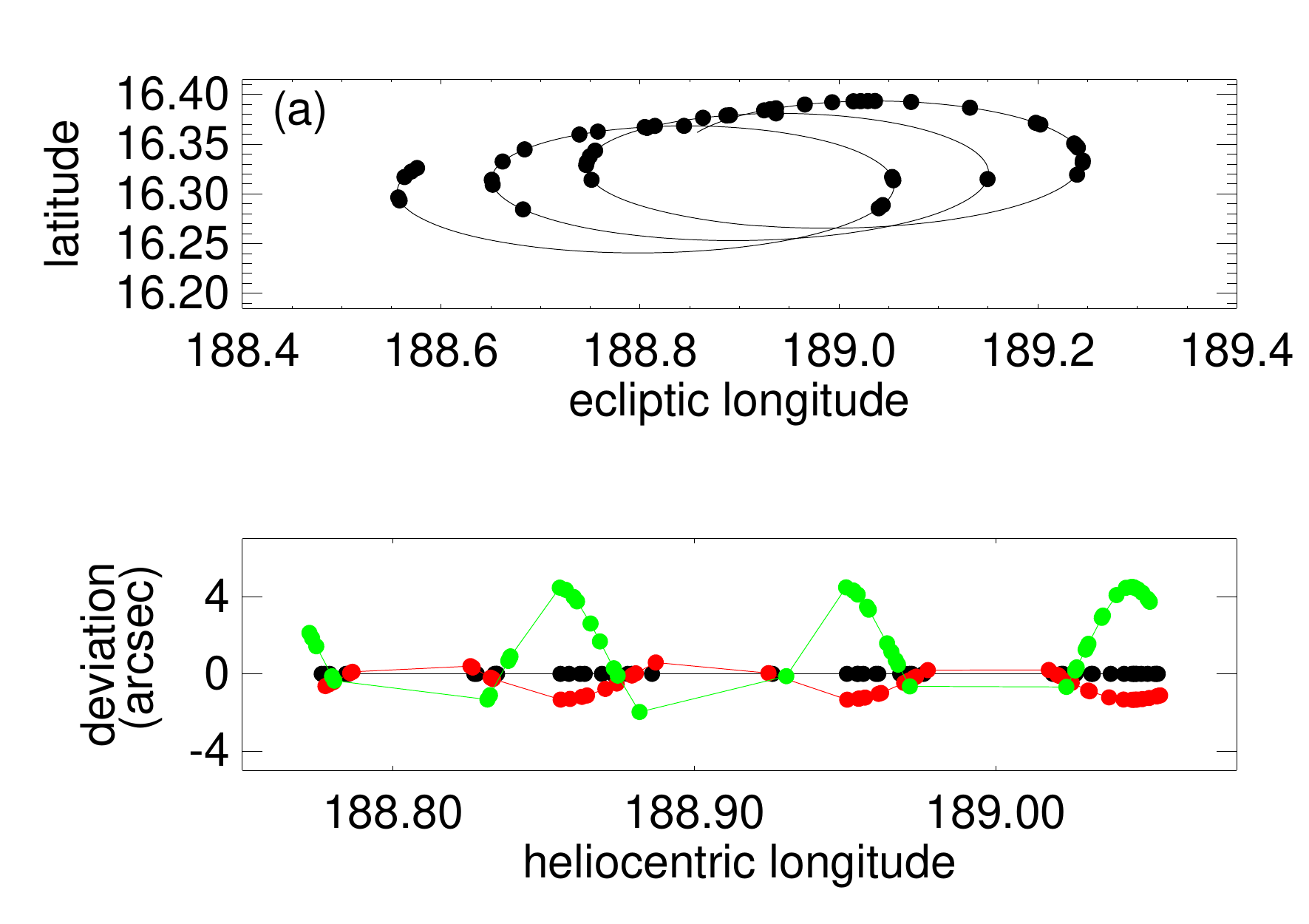}
   \caption{An example object from our Planet
   Nine reference population with a 
   current distance of 264.8 AU is
   predicted to have been detected 50 times in the ZTF
   survey. (a) Seen from the Earth, these detections make
   three loops over the three year survey period. 
   (b) Transformed to heliocentric coordinates at the
   correct assumed distance (black), the object travels
   in a straight line at a constant speed. With an
   incorrect assumed distance of 263.2 AU (red) or 270.3 AU (green)
   deviations to the straight line motion can be seen.}
\end{figure}

In our implementation, we assume a range for heliocentric distance, $r$, 
(following Holman et al., we explicitly assume uniform spacing in $\gamma=1/r$ from zero to
$1/r_{\rm min}$, the minimum distance considered, here set to 200 AU), and
then, for each detected transient and assumed distance, we calculate the
great circle direction and speed required to link every other single transient
in the catalog (in practice, we calculate $\dot\alpha$ and $\dot\beta$,
the speed in heliocentric longitude and in latitude). 
Real objects will appear as a cluster within a fixed range of
$\Delta\dot\alpha$ and $\Delta\dot\beta$ 
when the assumed distance is close to their correct distance, 
demonstrating that they are consistent with great circle
motion at a constant speed (Figure 3). Depending on our assumed tolerances,
sometimes random collections of false-positive transients will
also appear sufficiently clustered. 
To further filter out such false linkages, we take each 
potential cluster, perform an iterative calculation for the best fit
value of $r$ using the heliocentric Holman approximation, and calculate RMS
astrometric residuals, RMS$_{\rm holman}$. All clusters which pass this
RMS threshold are
passed to a full orbit integrator using the Bernstein
geocentric algorithm and a final astrometric residual RMS$_{\rm bernstein}$ threshold is imposed.
In regions of high space density of transients,
the number of transients, $m$ within a $\Delta\dot\alpha$, $\Delta\dot\beta$
cluster threshold may
exceed the number, $n$, required for a detection. In these
cases we create and individually test all $_mC_n$ combinations
containing $n$ transients independently.
\begin{figure}
\epsscale{1.25}
  \plotone{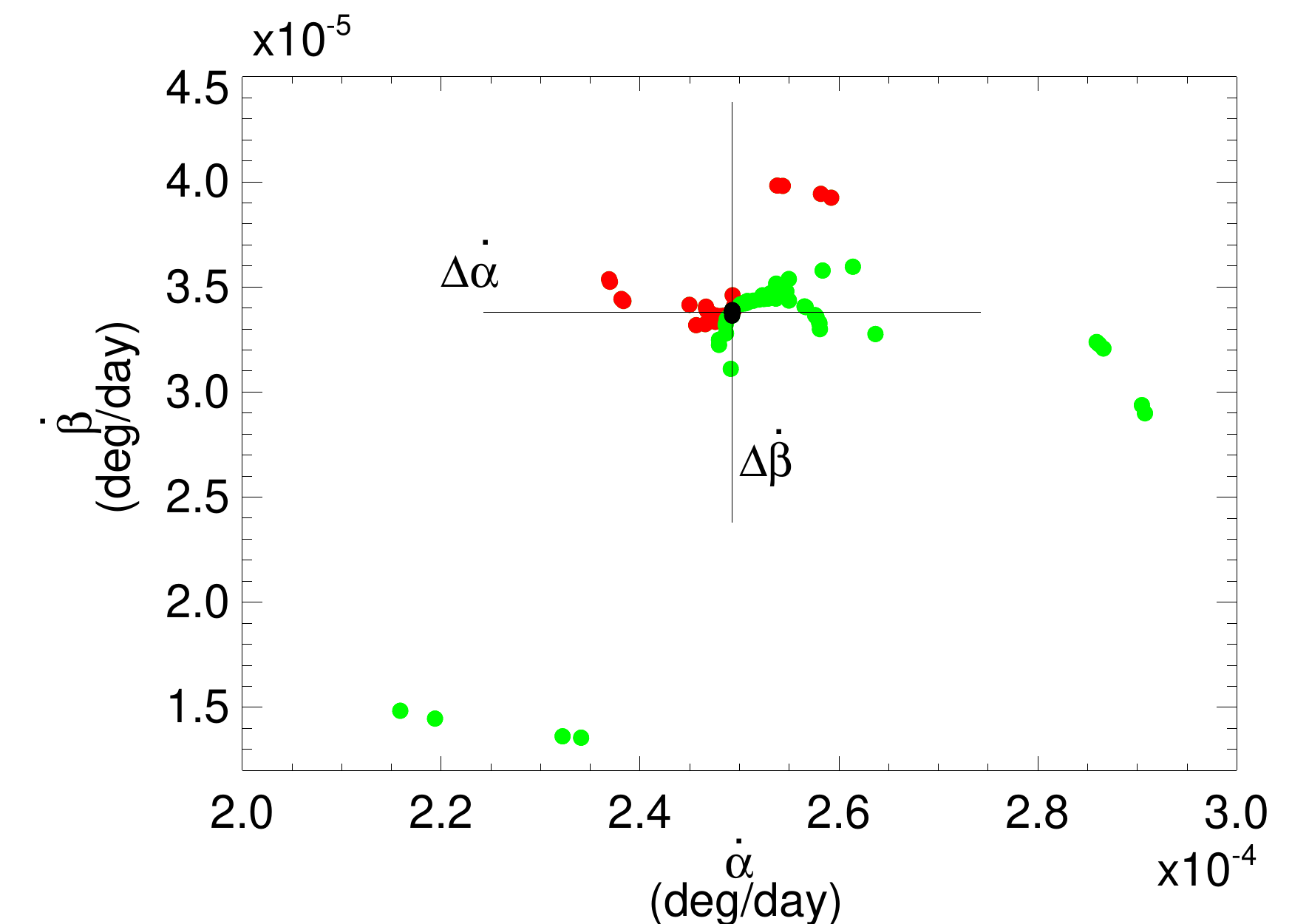}
  \caption{The speed in heliocentric longitude and in 
  heliocentric latitude to travel from the first point
  in the orbit to each of the other 49 for our 
  synthetic object shown in Fig. 2. If the correct
  distance of 264.8 AU is assumed, the speeds to all data points are
  nearly identical (black). With an incorrect assumed
  distance (263.2 AU, red, and 270.3 AU, green, as in 
  Figure 2), the cluster of inferred speeds is larger. Our algorithm
  finds all clusters within the ranges of selected
  values of $\Delta\dot\alpha$ and $\Delta\dot\beta$
  shown and would successfully identify all red (and black)
  detections as clustered. With the incorrect
  distance assumed for the green cluster, many
  of the points would not be identified.}
\end{figure}

In this method, the five critical parameters to be chosen,
each of which affect algorithmic efficiency, are $n$,
the number of detections required,
the grid spacing of assumed heliocentric distances used, $\Delta\gamma$,
the tolerances in heliocentric speed in longitude and latitude, $\Delta\dot\alpha$
and $\Delta\dot\beta$,
and the two RMS
astrometric residual thresholds, RMS$_{\rm holman}$ and RMS$_{\rm bernstein}$.
The initial step of heliocentric projection and speed
calculation involves computationally
efficient matrix operations. 
Each potential cluster that needs to be passed to the iterative
calculation for best-fit $r$,
however, requires significantly longer computational time, so we attempt to
limit the number of these calculations.
Fewer false clusters will be found if $n$, the number of required
detections, is larger, but then the probability of 
finding real objects decreases. Smaller numbers of
false clusters can also be found if $\Delta\dot\alpha$ and $\Delta\dot\beta$
are made tighter, however to do so requires an increasingly
finer grid in assumed distances, $\Delta\gamma$, as even real clusters 
appear dispersed when projected to a sufficiently 
wrong heliocentric
distance. The cluster dispersion problem becomes more
acute for objects at smaller heliocentric distances and
for observations over larger time spans, both of which
begin to stress the assumptions of the Holman et al. method. 

We determine efficient values of these parameters by considering the
simulated observations of the Planet Nine reference population. First,
for computational efficiency and ease of parallelization, we
divide the sky into three-by-three-degree bricks to 
be processed separately (each brick overlaps adjacent bricks
sufficiently that all detections of
the fastest-moving object expected will appear within at least
one single brick, as verified by our reference population). For each brick, we collect all simulated
reference population detections and test our linking algorithm 
multiple times using a logarithmically-spaced choice for $\Delta\gamma$ and,
for each, calculating the minimum
values of $\Delta\dot\alpha$ and $\Delta\dot\beta$ 
required to link all observations.
We then calculate the astrometric residuals after an
iterative heliocentric best fit to $r$ and save the maximum
value obtained in each brick as RMS$_{\rm holman}$. We also solve
each object using the full geocentric integration and likewise
save the maximum astrometric residual as RMS$_{\rm bernstein}$. 
{ We add 1 arcsecond to each of these RMS parameters to account
for the astrometric uncertainties of the real observations.}
As a computational convenience, for each
brick we also calculate upper and lower limits to the speed
and direction for all members of the Planet Nine reference population. 
In our final analysis
we will restrict ourselves to only searching for links that
are within these limits.
At this stage we have appropriate $\Delta\dot\alpha$ and $\Delta\dot\beta$
for a wide selection of potential choices of $\Delta\gamma$.
The final choice
of $\Delta\gamma$ and required numbers of detections
comes from choosing the sparsest grid possible and minimum
number of detections
that do not cause too many initial linkages that need
to be checked with the iterative algorithms. In practice we
make this choice by examining our most densely populated brick
and running a few test cases. For this analysis we find
that requiring 7 detections and a grid spacing of $\Delta\gamma$=0.0001 AU$^{-1}$ leads
to efficient processing. With this choice for $\Delta\gamma$, typical values
for $\Delta\dot\alpha$ and $\Delta\dot\beta$ are $5\times 10^{-5}$ and $2\times 10^{-5}$ deg/day,
respectively, with clear dependence 
on ecliptic latitude. Values for RMS$_{\rm holman}$ range from 2 to 12 arcseconds (again,
with the major dependence on ecliptic latitude), and a global value for RMS$_{\rm bernstein}=1$ arcsec
is found to be appropriate, which we increase to RMS$_{\rm bernstein}$=2 arcsec to account for 
astrometric uncertainties not present in the synthetic reference population data.

An example of the workings of the algorithm
can be seen in Figures 2 and 3.
  One of the members of the Planet Nine reference
  population, with V=19.9 and at a 
   current distance of 264.8 AU, is
   predicted to have been detectable 50 times in the ZTF
   survey. Seen from the Earth, these detections
   make 
   three loops over the three year survey period 
   (Fig 2a). Efficiently linking these observations 
   within a background of randomly placed false
   positive detections
   is a difficult problem. 
   When transformed to heliocentric coordinates at the
   correct distance (black point in Fig 2b), 
   the object travels in a straight line at a constant
   speed. With our choice of $\Delta\gamma=0.0001$ AU$^{-1}$, the closest assumed distance to the true
   distance is 263.2 AU (with the next closest 270.3 AU).
   Using those values as our assumed distances (red and green respectively), divergences from the straight line
   motion begin to appear.
   Figure 3 shows the speed in heliocentric longitude
   and latitude required for an object to move from the
   first position of the orbit to each of the 49 other 
   positions of the orbit. If the correct distance is
   assumed, these speeds are all nearly identical (black
   points), demonstrating constant circular motion. At the closest distance in our assumed 
   grid, all of the speeds still fall within the
   range of $\Delta\dot\alpha$ and $\Delta\dot\beta$
   such that they would be flagged as a cluster to
   further study. In this particular case, the iterative
   Holman fit finds a distance of 264.8 AU and an RMS
   of 0.10 arcseconds. The full Bernstein integration
   finds an RMS of 0.30 arcseconds. All 50 detections
   of this object would be successfully linked across
   three oppositions.

After selection of algorithm parameters, each brick is ready to be searched
for Planet Nine. Before proceeding, we inject the entire data set of simulated
observations of the Planet Nine reference population,
adding normally distributed errors of 1 arcsecond to
each astrometric position to conservatively
account for the measurement uncertainties in ZTF. 
While these objects should all 
be detected by design, embedding them within the real data allows an end-to-end
calibration of our search method. Each of the 1164 bricks is processed
independently. Of the 56373 members of the Planet Nine reference population
that had 7 or more detections that
we injected into the data, { 56173, or 99.66\%}
were detected and linked. Our algorithm is extremely
efficient at successfully linking detections over the full three
year period for these distant objects.

In the full ZTF data set,
103 potential linkages made it to the final
step of full orbital integration using the Bernstein integration
method -- including one with as many as 23 linkages -- and all but one failed
this final cut. 
The single object that survived the final cut was, surprisingly, Eris.
Eris is coincidentally along the predicted P9 path, but the
search algorithm was not optimized to find objects
at a distance of only 96 AU. Nonetheless, Eris was successfully
linked owing to a string of 11 detections in 20 days that
occurred close to
quadrature when the motion of Eris could not be distinguished
from that of a more distant object.
We examined each of the 102 other linkages
individually and did not find any 
reason to consider them further. Planet Nine was not detected in the ZTF data.

\begin{figure}
   \plotone{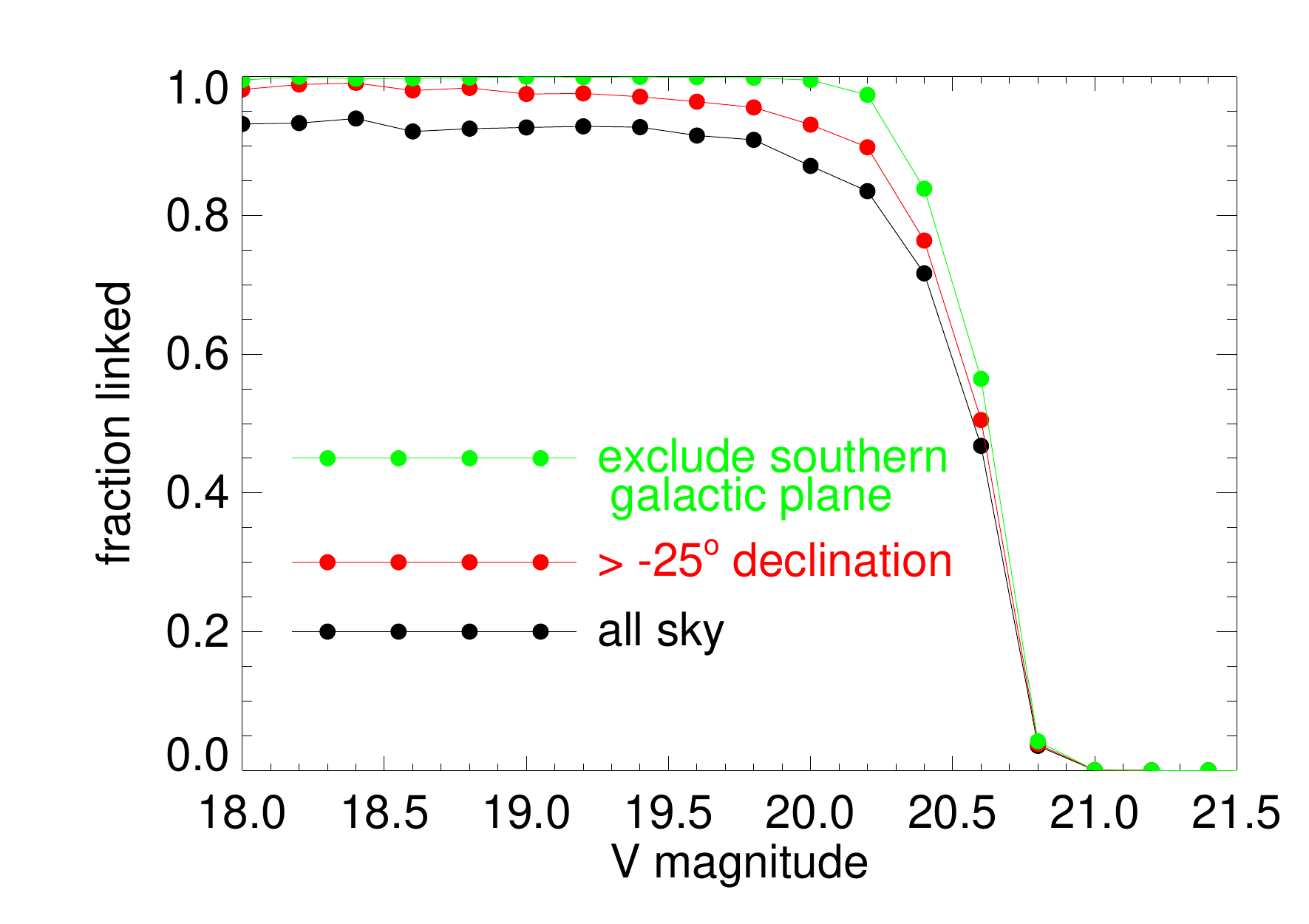}
   \caption{Limits to the detection of the Planet Nine reference population
   in the ZTF data. Across the entire sky 91\% of the brightest objects are detected. If we exclude areas
   south of -25$^{\circ}$ declination, 94\% are detected,
   and if we exclude the poorly observed southern
   hemisphere galactic plane region, 100\% of the bright
   objects are detected. }
\end{figure}
\begin{figure*}
\plotone{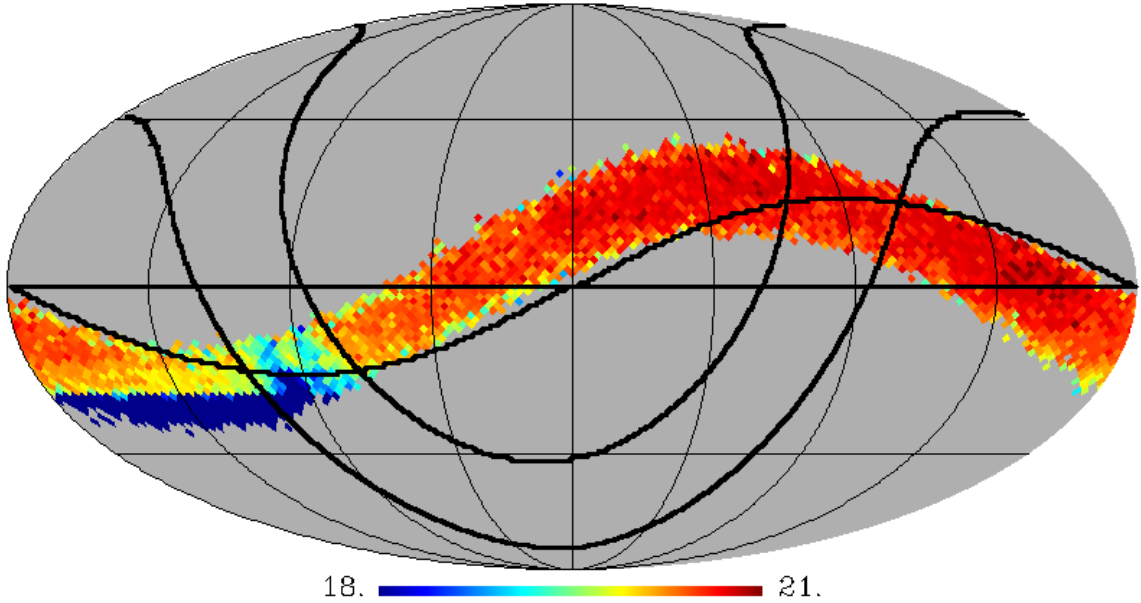}
\caption{Detection limits across the Planet Nine search region determined from the search for objects from the Planet Nine reference population injected
into the data stream.
The geometry of the project is identical to that of Figure 1. The dark blue regions
in the southern hemisphere are areas not covered by
ZTF.}

\end{figure*}
\section{Planet Nine detection limits}
With no detection of Planet Nine in the data set we 
use the Planet Nine reference population to determine limits across
the search region. Our simplest metric is the
fraction of the 100,000 members of the Planet Nine reference population
that { would have been detected. Our algorithm 
would have found 56173
members of the reference population in ZTF across the full sky. The brightest
members of the population are most efficiently detected, as shown
in Figure 4.
The ZTF survey would have
successfully identified $\sim$92.4\% of all members
of the Planet Nine reference population with $V<19.75$},
with a 50\% efficiency at $V=20.57$.
Many of the bright objects that were not linked were further south than the declination
limit of the ZTF survey. If we examine the fraction of objects successfully linked that have
declinations above -25$^\circ$, we find the the survey linked 95\% of the bright objects, with 
a 50\% efficiency at $V=20.60$. Of the bright objects within the ZTF survey region, the other 
source of missed detections is the poorer coverage in the southern galactic plane. If we exclude
the southern regions with galactic latitude less than 15$^\circ$ we detect 99\% of the objects brighter
than $V=20.0$ and a 50\% efficiency at $V=20.62$.

To better examine the spatial structure of the detection limits, we again use our synthetic population,
and we determine the brightest synthetic Planet Nine that is missed at each HEALPix. We define the limit
as the next-brightest object from this first missed object. A map of these limits is show in Figure 5,
which also conveniently shows the Planet Nine search region. For the Planet Nine search region, the limits
found from the Planet Nine reference population are nearly identical to those of Figure 1 determined for asteroids.
The asteroid limits, requiring a 95\% detection efficiency, are better sampled and better
defined than the synthetic  Planet Nine limits, so we regard the asteroid limits of Fig. 1 the
magnitude limits of the full survey. 
\section{Discussion}

The results presented here provide a useful framework for understanding 
our current limits on the detection of Planet Nine and for incorporating 
additional surveys into the limits. Any survey can use the 
the full reference population of 100,000 synthetic Planets Nine 
to inject samples from the BB21 analysis and determine if
they are detected. Each reference object
contains full orbital elements as well as 
mass, assumed radius, and assumed albedo, 
and can be used to calculate positions, distances, magnitudes, thermal fluxes, 
and gravitational forces at any survey time. Different
assumptions about radius and albedo can easily be incorporated
into the population.
The full reference population is
hosted at \url{https://data.caltech.edu/records/2098} \citep{brown_2021}, which also indicates which objects from this population would
have been detected in the ZTF survey. The table of potential detections
will be expanded as additional surveys use this reference population
and determine which objects they should have detected\footnote{Contact 
the first author to include analysis from a survey into the reference
population tables.}.
The reference population as well as the list of objects which should have
been detected
can easily be used to recreate Figures 4 and 5 for the current survey
and equivalent visualizations for all surveys that are included.

The website also hosts full HEALPix limit map equivalent to Figure 5
and will provide updates
for any additional surveys which publish equivalent limits. 
This map will be able to be used to quickly discern
which areas of the sky have had published surveys contributing to the global limits on Planet Nine, with the 
important caveat that the limits presented here, at least, are only for objects with roughly the
orbital characteristics predicted for Planet Nine by BB21. 

These are the first 
rigorous optical limits placed on the existence of Planet Nine over nearly
the full predicted search area. This survey could have
detected more than 56\% of the P9 reference population. 
Given the assumptions of radius and albedo that we
used for the reference population, we can update expected
Planet Nine parameters. The remaining population has 15.9, 50.0, and 84.1 percentile values for mass of $6.3_{-1.5}^{+2.3}$ Earth masses,
semimajor axis of $460_{-100}^{+160}$ AU, perihelion of $340_{-70}^{+80}$ AU and aphelion of $560_{-140}^{+260}$ AU.
The other parameters are generally unchanged. Not surprisingly,
a survey ruling out more than half of the most bright parameter
space pushes the expected orbit of P9 further away.

The method that we have developed here for
both detection and calibration of transient surveys is applicable 
to many other existing surveys. Given the potential of a brighter 
Planet Nine than original anticipated, such searches could prove fruitful.

\acknowledgements
This research has benefited greatly from conversations with 
Rich Dekany, Matt Holman, Matt Payne, and Matthew Belyakov. An anonymous
referee provided insightful comments that improved
the presentation of this manuscript.

\bibliography{kbo0421}{}
\bibliographystyle{aasjournal}

\end{document}